\documentclass{PoS}
\usepackage{amssymb,amsmath,bm}
\DeclareMathOperator{\tr}{Tr}
\renewcommand{\d}{\mathrm{d}}
\newcommand{\sT}{{\scriptscriptstyle T}}

\title{Linear polarization of gluons and Higgs plus jet production at the LHC}

\ShortTitle{Linear polarization of gluons and Higgs plus jet production at the LHC}

\author{\speaker{Cristian PISANO}\\
    Department of Physics, University of Antwerp, Groenenborgerlaan 171, 2020 Antwerp, Belgium\\
      E-mail: \email{cristian.pisano@uantwerp.be}}

\abstract{We show that the production of a  Higgs boson in association with a jet in proton-proton collisions is sensitive to the linear polarization of gluons inside unpolarized protons.  We present the analytical expressions, at leading order in perturbative QCD, for various transverse momentum dependent observables, which can be measured at the LHC. Since there are no experimental constraints on the linearly polarized gluon distribution, its effects are studied by adopting two different models. In particular, we find that the $\cos2\phi$ azimuthal asymmetry is a very promising observable because it could give access to the sign of this new distribution.}

\FullConference{XXIII International Workshop on Deep-Inelastic Scattering\\

                 27 April - May 1 2015\\

                 Dallas, Texas}

\begin{document}

\section{Introduction}

Gluons can be linearly polarized, even inside an unpolarized hadron, if one takes into account their transverse momentum with respect to the hadron momentum~\cite{Mulders:2000sh}.  The amount of polarization, corresponding to an interference between $+1$ and $-1$ gluon helicity states, is so far unknown. However, if sufficiently large it could provide, for instance, a valuable new tool to analyze the couplings of the Higgs boson to the standard model particles into which it can decay~\cite{Boer:2013fca}. The effects of gluon polarization can be studied in a convenient way adopting the formalism of transverse momentum dependent (TMD) parton distribution functions (TMDs, for short). Within this framework, inclusive Higgs production has been studied in Refs.~\cite{Boer:2013fca,Boer:2011kf} and, including the effects of TMD evolution, in Refs.~\cite{Boer:2014tka,Echevarria:2015uaa}.  
The impact of gluon polarization is expected to be small, most likely a few percent, at the energy scale of the Higgs mass $M_H$. Moreover, these effects are largest when the transverse momentum of the Higgs boson is small, i.e.\ a few GeV, where the cross section is difficult to measure.  

Alternatively, in the following, we propose the study of Higgs production in association 
with a jet~\cite{Boer:2014lka}. When it comes to probe the linear polarization of gluons, this process offers some additional features compared to inclusive Higgs production. First of all, one can study the TMD evolution by tuning the hard scale, identified for example with the invariant mass of
 the Higgs-jet pair.
This is not possible in Higgs production, when the hard scale is fixed to be $M_H$. Furthermore, 
it is possible to define angular modulations, which have the advantage of singling out specific contributions. Finally, we note that the effects of gluon polarization show up in the transverse momentum distribution of the Higgs-jet pair, where the transverse momentum of the pair can be as small as a few GeV, but the single transverse momenta of the Higgs boson and the jet might be much larger.

\section{Theoretical framework}
We consider the process
$p(P_A)\,{+}\,p(P_B)\,\to\, H (K_H) \,{+}\, {\rm jet} (K_{\rm j})\, {+}  \,X$,
where the four-momenta of the particles are given within brackets, and the Higgs boson and the jet are almost back to back in the plane perpendicular to the direction of the incoming protons. At leading
order in perturbative QCD, the partonic subprocesses that contribute are $gg\to Hg$, $gq\to Hq$ and 
$q \bar{q} \to Hg$.  We perform a Sudakov decomposition of the initial hadronic momenta $P_A$ and $P_B$ in terms of the lightlike vectors $n_+$ and $n_-$, with $n_+\cdot n_-=1$: 
\begin{equation}
P_A^\mu
=P_A^+n_+^\mu+\frac{M_p^2}{2P_A^+}n_-^\mu\ ,\qquad
P_B^\mu
=\frac{M_p^2}{2P_B^-}n_+^\mu+P_B^-n_-^\mu\,,
\end{equation}
where $M_p$ is the proton mass. Similarly, one can express the momenta of the two incoming partons, $p_a$ and $p_b$,  as follows
\begin{equation}
p_a^\mu
=x_a^{\phantom{+}}\!P_A^+n_+^\mu
+\frac{p_a^2{+}\boldsymbol p_{a \sT}^2}{2\,x_a^{\phantom{+}}\!P_A^+}n_-^\mu
+p_{a \sT}^\mu\ ,
\qquad p_b^\mu
=\frac{p_b^2{+}\boldsymbol p_{b \sT}^2}{2\,x_b^{\phantom{-}}\!P_B^-}n_+^\mu
+x_b^{\phantom{-}}\!P_B^-n_-^\mu+p_{b \sT}^\mu\, ,
\label{PartonDecompositions}
\end{equation}
with $x_a$, $x_b$ being the partonic light cone momentum fractions and $p_{a\sT}$,  $p_{b\sT}$ the intrinsic transverse momenta. Moreover, we define the difference and sum of the final transverse 
momenta, $\bm K_\perp = (\bm K_{H \perp} - \bm K_{{\rm j} \perp})/2$ and 
$\bm q_\sT = \bm K_{H \perp} + \bm K_{{\rm j} \perp}$, with $\vert \bm q_\sT\vert \ll \vert \bm K_\perp\vert $ because the Higgs and jet transverse momenta are almost opposite. The azimuthal angle between  $\bm K_\perp$ and $\bm q_\sT$ is denoted by $ \phi \equiv  \phi_\perp- \phi_\sT$.

Restricting ourselves to the channel $gg\to Hg$, which is the dominant one at the LHC energies, and assuming TMD factorization, the cross section for the process $pp\to H\,\text{jet}\,X$ is given by
\begin{eqnarray}
\d\sigma
& = &\frac{1}{2 s}\,\frac{d^3 \bm K_H}{(2\pi)^3\,2 E_{H}}\, \frac{d^3 \bm K_{\rm j}}{(2\pi)^3\,2 E_{\rm j}}\,
{\int} \d x_a \,\d x_b \,\d^2\bm p_{a\sT} \,\d^2\bm p_{b\sT}\,(2\pi)^4
\delta^4(p_a{+} p_b {-}K_H- K_{\rm j})
 \nonumber \\
&&\qquad \qquad\qquad \times
{\rm Tr}\, \left \{ \Phi^{[U]}_g(x_a {,}\bm p_{a \sT}) \Phi^{[U]}_g(x_b {,}\bm p_{b \sT})
 \left|{\cal M}^{g g   \to H g} (p_a, p_b; K_H, K_{\rm j})\right|^2\right \}\,,
\label{CrossSec}
\end{eqnarray}
where $s=(P_A+P_B)^2$, the trace is taken over the Lorentz indices and ${\cal M}^{gg\to Hg}$ is the 
amplitude for the partonic subprocess. The correlators $\Phi^{[U]}_g$, describing the proton $\to$ 
gluon transitions, are defined in terms of matrix elements of the gluon field strengths  $F^{\mu\nu}(0)$ and $F^{\mu\nu}(\xi)$  on the light front (LF) $\xi\cdot n\equiv 0$~\cite{Mulders:2000sh}, 
\begin{eqnarray}
\label{GluonCorr}
\Phi_g^{[U]\,\mu\nu}(x,\bm p_\sT )
& = &  \frac{n_\rho\,n_\sigma}{(p{\cdot}n)^2}
{\int}\frac{\d(\xi{\cdot}P)\,\d^2\xi_\sT}{(2\pi)^3}\
e^{ip\cdot\xi}\,
\langle P|\,\tr\big[\,F^{\mu\rho}(0)\,U_{[0,\xi]}\,
F^{\nu\sigma}(\xi)\,U^{\prime}_{[\xi,0]}\,\big]
\,|P \rangle\,\big\rfloor_{\text{LF}} \nonumber \\
& =& 
-\frac{1}{2x}\,\bigg \{g_\sT^{\mu\nu}\,f_1^g (x,\bm p_\sT^2)
-\bigg(\frac{p_\sT^\mu p_\sT^\nu}{M_p^2}\,
{+}\,g_\sT^{\mu\nu}\frac{\bm p_\sT^2}{2M_p^2}\bigg)
\;h_1^{\perp\,g} (x,\bm p_\sT^2) \bigg \} , \label{Phig}
\label{eq:corrg}
\end{eqnarray}
with $n\equiv n_-$ for gluon $a$ with momentum $p_a$ and $n\equiv n_+$ for gluon $b$ with momentum $p_b$. The process dependent gauge links $U_{[0,\xi]}$ and $U^{\prime}_{[\xi,0]}$ are needed to render the correlator gauge invariant, while the transverse tensor $g^{\mu\nu}_\sT$ is 
defined as $g^{\mu\nu}_{\sT} = g^{\mu\nu} - n_+^{\mu}n_-^{\nu} -n_-^{\mu}n_+^{\nu}$. 
The unpolarized and linearly polarized gluon TMD distributions are denoted respectively by 
$f_1^g (x,\bm p_\sT^2)$ and $h_1^{\perp\,g} (x,\bm p_\sT^2)$, and satisfy the model-independent relation~\cite{Mulders:2000sh}
\begin{equation}
\frac{\bm p_\sT^2}{2M_p^2}\,\vert h_1^{\perp \,g}(x,\bm p_\sT^2) \vert \le  f_1^g(x,\bm p_\sT^2)~.
\label{eq:bound}
\end{equation}
Since  $h_1^{\perp\,g}$ is $T$-even, it can be nonzero in absence of initial and/or final state interactions. However, such interactions can render it process dependent, like all other TMDs, or even lead to the breaking of factorization. Here we will not consider such effects.
We just note that there are at present no indications that factorization may be broken in $pp\to H\,\text{jet}\,X$  because of color entanglement as in $pp\to \text{jet}\,\text{jet}\,X$~\cite{Rogers:2010dm}.

From Eqs.~(\ref{CrossSec}), (\ref{eq:corrg}) and the explicit expression for the amplitude 
${\cal M}^{gg\to Hg}$, one can calculate the normalized cross section for the process $p\,p\to H\,{\rm jet}\,X$, defined as 
\begin{equation}
\frac{\d\sigma}{ \sigma} \equiv
\frac{ \d\sigma}
{\int_0^{q_{\sT \rm max }^2} \d \bm q_\sT^2\int_0^{2 \pi} \d\phi\, \d\sigma}\,, \qquad \text{with}\qquad
\d\sigma \equiv  \frac{\d\sigma}{\d y_H\, \d y_{\rm j} \,\d^2 \bm{K}_{\perp} \,\d^2 \bm q_\sT} \,,
\label{eq:qTdist}
\end{equation}
where $y_H$ and $y_j$ are, respectively, the rapidities of the produced Higgs boson and jet along the 
direction of the incoming protons. By neglecting  terms suppressed by powers of $\vert \bm q_\sT \vert /  M_\perp $, with $M_\perp = \sqrt{M_H^2 + \bm K_{H \perp}^2}$, the final result in the laboratory frame takes the form
\begin{eqnarray}
\frac{\d \sigma}{\sigma} & = & \frac{1}{2\pi}\,\sigma_0(\bm q_\sT^2) \,\left [ 1 + R_0(\bm q_\sT^2) + R_2(\bm q_\sT^2) \cos2\phi +  R_4(\bm q_\sT^2) \cos4\phi\right ]\,,
\label{eq:csgg-2}
\end{eqnarray}
where
\begin{equation}
\sigma_0(\bm q_\sT^2) \equiv \frac{ {\cal C}[f_1^g \, f_1^g ]}{\int_0^{{q^2_{\sT \rm max }}} \d \bm q^2_\sT \, {\cal C}[f_1^g \, f_1^g ] }\,,
\label{eq:sigma0}
\end{equation}
and we have introduced the convolution of TMDs
\begin{eqnarray}
{\cal{C}}[w\, f\, g] & \equiv & \int d^{2}\bm p_{a\sT}\int d^{2}\bm p_{b\sT}\,
\delta^{2}(\bm p_{a\sT}+\bm p_{b\sT}-\bm q_{\sT}) \, w(\bm p_{a\sT},\bm p_{b\sT})\, f(x_{a},\bm p_{a\sT}^{2})\, g(x_{b},\bm p_{b\sT}^{2})\,,
\label{eq:Conv}
\end{eqnarray} 
with,  up to corrections of order ${\cal O}(1/s)$,
$x_{a/b} \, =\, \left (M_{\perp}\,e^{\pm y_H} \, + \, \vert \bm K_{{\rm j} \perp} \vert \,
e^{\pm y_{\rm j}} \right )/{\sqrt s}$. 
The different terms $R_0$, $R_2$ and $R_4$ are functions of the Mandelstam variables 
$\hat s$, $\hat t$, $\hat u$ for the subprocess $gg\to Hg$ and contain convolutions of the gluon TMDs.
They read:
\begin{eqnarray}
R_0(\bm q_\sT^2)& =& \frac{M_H^4\,\hat s^2}{M_H^8 + \hat s^4 + \hat t^4 + \hat u^4}\, \frac{{\cal C}[w_0^{hh}\,h_1^{\perp\,g} \, h_1^{\perp\,g} ] } {{\cal C}[f_1^g \, f_1^g ]}\,, \label{eq:R0}\nonumber \\
R_2(\bm q_\sT^2)& = & \frac{ \hat t ^2(\hat t + \hat u)^2 -2 M_H^2 \hat u^2 (\hat t + \hat u) + M_H^4 (\hat t^2 + \hat u^2)  }{M_H^8 + \hat s^4 + \hat t^4 + 
\hat u^4 }  \, \frac{{\cal C}[w_2^{fh} \,f_1^g\,h_1^{\perp\,g}]}{{\cal C}[f_1^g \, f_1^g ]}  \,+ \,(x_a \leftrightarrow x_b , \,\hat t \leftrightarrow \hat u )\,,
\label{eq:R2}  \nonumber  \\
R_4(\bm q_\sT^2) & = & \frac{\hat t^2 \hat u^2}{M_H^8+\hat s^4 + \hat t^4 + \hat u^4} \,\frac{{\cal C}[w_4^{hh}\,h_1^{\perp\,g}h_1^{\perp\,g}]}{{\cal C}[f_1^g \, f_1^g ]}\,,
\label{eq:R4}
\end{eqnarray}
with the transverse weights given by
\begin{eqnarray}
w_0^{hh}  & = & \frac{1}{M_p^4}\, \left[ (\bm p_{a\sT}\cdot \bm p_{b\sT})^2 - \frac{1}{2}\, \bm p_{a\sT}^2 \,\bm p_{b\sT}^2\right ]\, ,\nonumber \\
w_2^{fh} & = & \frac{1}{M_p^2}\, \left [ 2\, 
\frac{(\bm q_\sT \cdot \bm p_{b\sT})^2}{\bm q_\sT^2} -\bm p_{b\sT}^2   \right ] \,,\quad
 w_2^{hf}\,  =  \, \frac{1}{M_p^2}\,\left [ 2\, 
\frac{(\bm q_\sT \cdot \bm p_{a\sT})^2}{\bm q_\sT^2}-\bm p_{a\sT}^2   \right ]\,,\nonumber \\
w_4^{hh} & = & \frac{1}{2 M_p^4}\, \left \{ 2 \,\left [
 2 \, \frac{(\bm q_\sT\cdot \bm p_{a\sT}) (\bm q_\sT\cdot \bm p_{b\sT})   }{\bm q_\sT^2} 
  -\bm p_{a\sT} \cdot  \bm p_{b\sT} \right ]^2 - \bm p_{a\sT}^2  \bm p_{b\sT}^2 \right \}~. 
\end{eqnarray}

In order to single out the different terms $1+R_0$, $R_2$, $R_4$ in 
Eq.~(\ref{eq:csgg-2}),  we define the observables~\cite{Dunnen:2014eta}
\begin{equation}
\langle \cos n\phi \rangle_{q_\sT} \equiv 
\frac{\int_0^{2 \pi} \d \phi \,\cos n\phi\, 
\d\sigma}{\sigma}\,,\qquad n=0, 2, 4\,,
\label{eq:cosnphiqT}
\end{equation}
such that their integrals over $\bm q_{\sT}^2$ give the average values of $\cos n\phi$. Namely,
\begin{equation}
\langle \cos n\phi \rangle \equiv \frac{\int_0^{q_{\sT \rm max }^2} \d\bm q_{\sT}^2\int_0^{2 \pi} \d \phi \,\cos n\phi\, 
\d\sigma}{\sigma}  = \int_0^{q_{\sT \rm max }^2} \d\bm q_{\sT}^2\,\langle \cos n\phi \rangle_{q_\sT} \,,\qquad n=0, 2, 4\,.
\label{eq:cosnphi}
\end{equation}
The choice of $q_{\sT \rm max }$ must be consistent with the requirement of TMD factorization that $q_\sT \ll Q$, with $Q$ being the hard scale of the process. Here we take $q_{\sT {\rm max}} =M_H/2$~\cite{Boer:2014lka}. It can be  shown that
\begin{eqnarray}
\frac{1}{\sigma}\,\frac{\d\sigma}{\d \bm q_{\sT}^2} & \equiv & \langle 1 \rangle_{q_\sT} \,  = \,  \sigma_0(\bm q_\sT^2)\, [ 1+ R_0(\bm q_\sT^2)]\,,
\label{eq:qT} \\
\langle \cos 2\phi \rangle_{q_\sT} & = &  \frac{1}{2}\,
\sigma_0(\bm q_{\sT}^2)\, R_2(\bm q_{\sT}^2)  \,, \\
\langle \cos 4\phi \rangle_{q_\sT} & = &   \frac{1}{2}\,
\sigma_0(\bm q_{\sT}^2)\, R_4(\bm q_{\sT}^2) \,.
\label{eq:cos4phi}
\end{eqnarray}

\section{Numerical results}
In this section we provide numerical estimates for the observables in Eqs.~(\ref{eq:qT})-(\ref{eq:cos4phi}), in the specific configuration in which the Higgs boson and the jet have the same rapidities ($y_H=y_{\rm j}$). To this aim, we assume that the so far unknown unpolarized gluon TMD 
has the following form~\cite{Boer:2014tka},
\begin{equation}
f_1^g(x,\bm p_\sT^2) = {f_1^g(x)}\, \frac{R^2}{2\,\pi}\, \frac{1}{1 + \bm p_\sT^2\,R^2}\,,
\label{eq:f1gdb}
\end{equation}
where  $R= 2$ GeV$^{-1}$ and $f_1^g(x)$ is the common gluon distribution, integrated over the transverse momentum. In order to show the maximal effects of gluon polarization, we take
$h_1^{\perp \,g}(x,\bm p_\sT^2)$ to be positive and saturating its positivity bound in Eq.~(\ref{eq:bound}),
\begin{equation}
 h_1^{\perp \,g}(x,\bm p_\sT^2)  = \frac{2M_p^2}{\bm p_\sT^2} f_1^g(x,\bm p_\sT^2)~.
\label{eq:h1pg}
\end{equation}
Alternatively, in analogy to Eq.~(\ref{eq:f1gdb}), we consider the following model as well, for which the bound is saturated only in the limit $p_\sT \to \infty$~\cite{Boer:2014tka}:
\begin{equation}
h_1^{\perp\,g}(x,\bm p_\sT^2) = c\,f_1^g(x)\, \frac{M^2_pR_h^4}{2\,\pi}\,
\frac{1}{(1+\bm p_\sT^2R_h^2)^2}\,,
\label{eq:h1pgdb}
\end{equation}
with $c= \pm 2$ and $R_h = 3 R/2$.  We note that there are no indications available on the actual shape of $h_1^{\perp\,g}$, therefore our models are only intended to illustrate what kind of features can arise qualitatively from linearly polarized gluons and to give an indication of the maximal effects that one might expect.  
\begin{figure}[t]
\centering
\includegraphics[width=6.4cm]{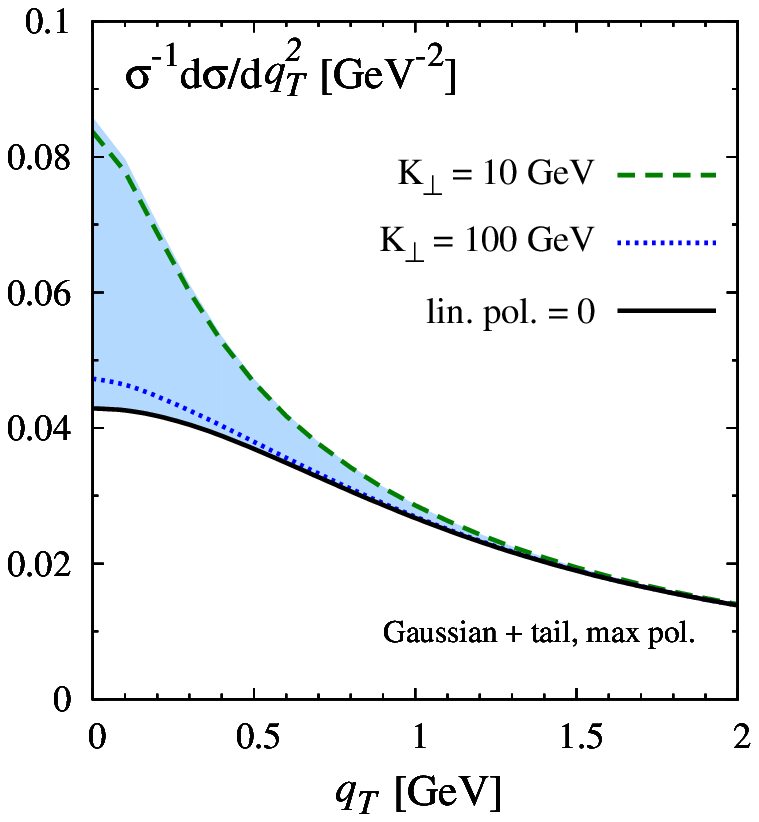}
\includegraphics[width=6.4cm]{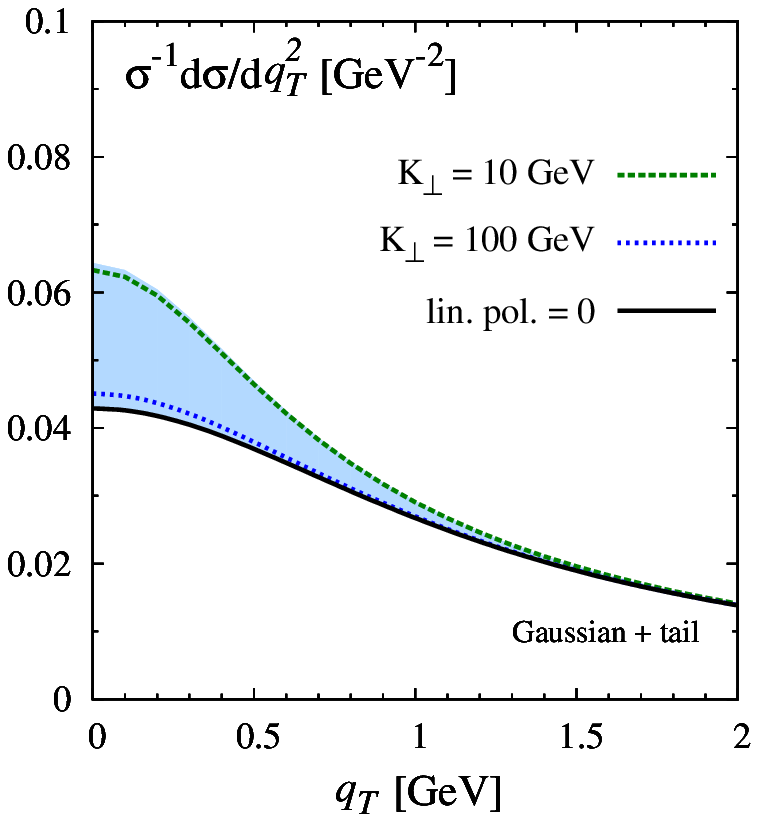}
\caption{Transverse momentum distribution of the Higgs boson plus jet pair in the process $p\,p\to H\,{\rm jet}\, X$ for two different choices of $K_\perp$, $K_\perp = 10 $ and 100 GeV, with $q_{\sT {\rm max}} =M_H/2$ and $y_H=y_{\rm j}$.  The solid line indicates the distribution in absence of linear polarization. The shaded blue area represents the range of the spectrum as $K_\perp$ varies from zero to infinity.}
\label{fig:qT_db}
\end{figure}

Our predictions for the transverse momentum distribution defined in Eq.~(\ref{eq:qT}) are presented in Fig.~\ref{fig:qT_db} for two different values of $K_\perp \equiv  \vert \bm K_\perp\vert $: 10 and 100 GeV. In the left panel of all the figures,  $h_1^{\perp\,g}$ is assumed to be maximal, see Eq.~(\ref{eq:h1pg}), while in the right one it is given by Eq.~(\ref{eq:h1pgdb}).

\begin{figure}[t]
\centering
\includegraphics[width=6.4cm]{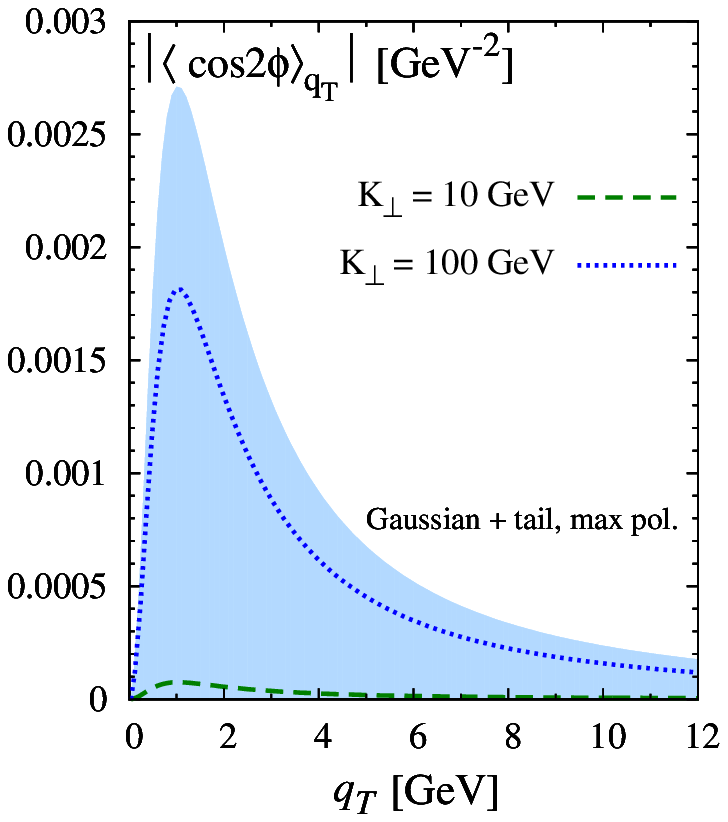}
\includegraphics[width=6.4cm]{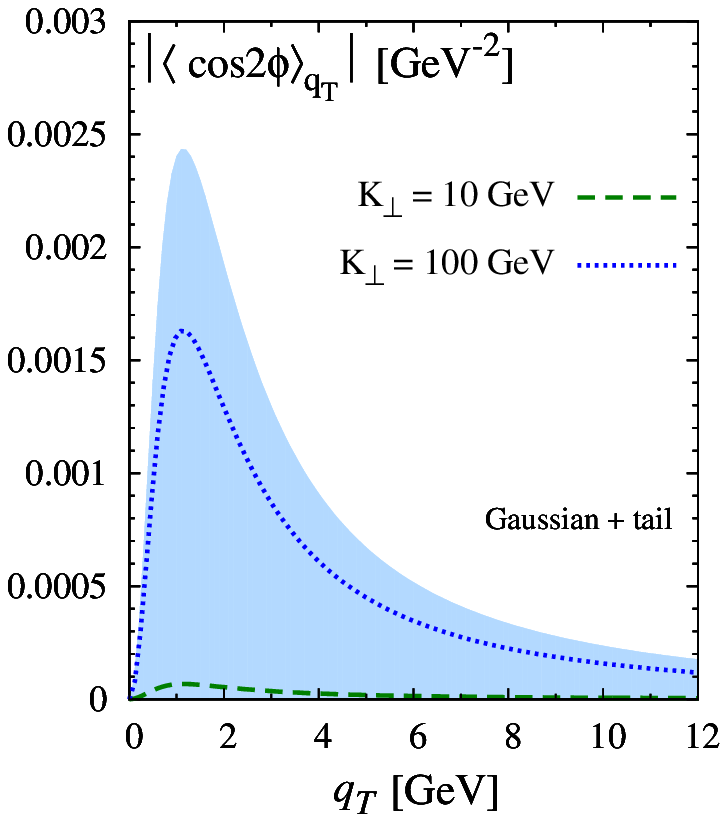}
\caption{Absolute value of the $ \langle \cos 2 \phi\rangle_{q_\sT}$  asymmetries the process $p\,p\to H\,{\rm jet}\, X$ for two different choices of $K_\perp$, $K_\perp = 10 $ and 100 GeV, with $q_{\sT {\rm max}} =M_H/2$ and $y_H=y_{\rm j}$. The shaded blue area represents the range of the asymmetries as $K_\perp$ varies from zero to infinity.}
\label{fig:cos2phi}
\end{figure}

\begin{figure}[b]
\centering
\includegraphics[width=6.4cm]{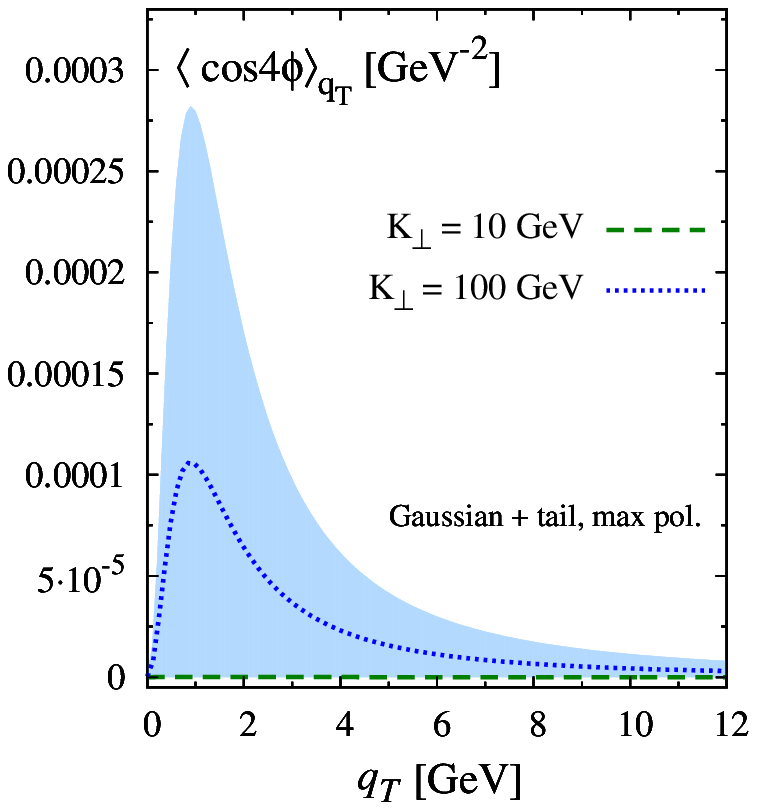}
\includegraphics[width=6.4cm]{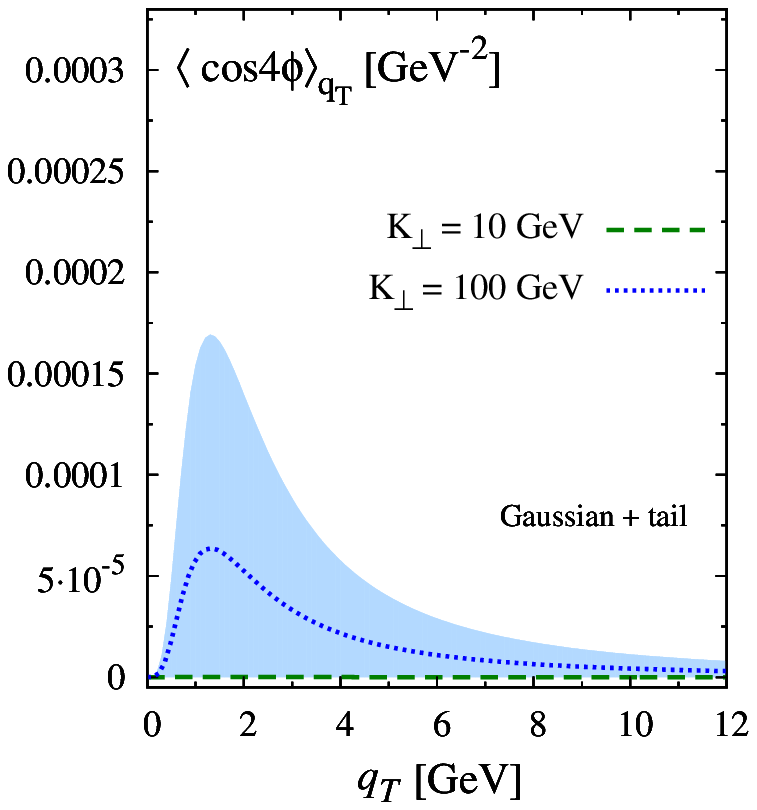}
\caption{Same as in Figure~2, but for the $ \langle \cos 4 \phi\rangle_{q_\sT}$  asymmetries.}
\label{fig:cos4phi}
\end{figure}

Among all the observables discussed here,  $\langle \cos 2 \phi\rangle_{q_\sT}$ is the only one which is sensitive to the sign of $h_1^{\perp\,g}$, and it is expected to be negative if $h_1^{\perp\,g} > 0$. Its absolute value as a function of $q_\sT$ is shown in Fig.~{\ref{fig:cos2phi}}. Furthermore, we find that $\vert\langle \cos 2 \phi\rangle\vert \approx 12\%$ if $K_\perp = 100$ GeV, while it is about 0.5\% if $K_\perp = 10$ GeV.  Analogously, our estimates for $\langle \cos 4 \phi\rangle_{q_\sT}$ are presented in Fig.~\ref{fig:cos4phi}, with $\langle \cos 4 \phi\rangle \approx 0.2 \%$ at $K_\perp = 100$ GeV and 
completely negligible at $K_\perp = 10$ GeV. Very similar values are obtained if one considers a Gaussian model for $f_1^g$ and maximal gluon polarization~\cite{Boer:2014lka}.

\section{Conclusions}
We have presented several observables for the process $pp\to H\, \text{jet}\,X$, which are sensitive 
to the polarized gluon TMD $h_1^{\perp\,g}$. We note that the proposed measurements 
are challenging, since one would need several bins in $q_\sT$ in the kinematic region up to about $10$ GeV. Therefore a high resolution of the transverse momentum of both the Higgs boson and the jet is required, in addition to the knowledge of how well the jet axis coincides with the direction of the fragmenting parton. 

To conclude, Higgs production at the LHC, both inclusive and in association with a jet,
complemented by  the study of processes likes $pp\to \eta_{c,b}\,X$, $pp\to \chi_{0 \,c,b}\,X$~\cite{Boer:2012bt} and $pp\to J/\psi\, \gamma\, X$~\cite{Dunnen:2014eta}, can be used to access $h_1^{\perp\,g}$ and to analyze its process and scale dependences. Additional information might be gathered  by looking at the electroproduction of heavy quark pairs and dijets~\cite{Boer:2010zf,Pisano:2013cya} that could be measured at  future EIC or LHeC experiments.

\acknowledgments
We acknowledge support by the Fonds Wetenschappelijk Onderzoek - Vlaanderen (FWO) through a Pegasus Marie Curie Fellowship.

\end{document}